\documentclass[acmsmall]{acmart}

\AtBeginDocument{%
  \providecommand\BibTeX{{%
    \normalfont B\kern-0.5em{\scshape i\kern-0.25em b}\kern-0.8em\TeX}}}

\acmConference[]{KDD Workshop on Humanitarian Mapping}{August 24, 2020}{accepted}
\acmBooktitle{KDD Workshop on Humanitarian Mapping, August 24, 2020}
\acmPrice{}
\acmISBN{}
\usepackage{multirow}
\usepackage{dblfloatfix}
\usepackage[linesnumbered,ruled]{algorithm2e}

\usepackage{color}
\begin{document}
\onecolumn 

\title[Downscaling Socioeconomic Census Data]{Magnify Your Population: Statistical Downscaling to Augment the Spatial Resolution of Socioeconomic Census Data}

\author{Giulia Carella}
\authornote{\textit{Corresponding author, \url{giulia@carto.com}, CARTO 307 5th Ave, New York, NY USA}}
\email{giulia@carto.com}
\orcid{0000-0001-7226-8231}
\affiliation{%
  \institution{CARTO}
  \streetaddress{46 Calle Gran Vía}
  \city{Madrid}
  \state{SPAIN}
  \postcode{}
}

\author{Andy Eschbacher}
\email{eschbacher@carto.com}
\affiliation{%
  \institution{CARTO}
  \streetaddress{4075 Wilson Boulevard}
  \city{Arlington}
  \state{VA, USA}
  \postcode{}
}

\author{Dongjie Fan}
\email{dfan@carto.com}
\affiliation{%
  \institution{CARTO}
  \streetaddress{307 5th Ave}
  \city{New York}
  \state{NY, USA}
  \postcode{}
}

\author{Miguel \'{A}lvarez}
\email{malvarez@carto.com}
\affiliation{%
  \institution{CARTO}
  \streetaddress{46 Calle Gran Vía}
  \city{Madrid}
  \state{SPAIN}
  \postcode{}
}

\author{\'{A}lvaro Arredondo}
\email{aarredondo@carto.com}
\affiliation{%
  \institution{CARTO}
  \streetaddress{46 Calle Gran Vía}
  \city{Madrid}
  \state{SPAIN}
  \postcode{}
}

\author{Alejandro Polvillo Hall}
\email{apolvillo@carto.com}
\affiliation{%
  \institution{CARTO}
  \streetaddress{46 Calle Gran Vía}
  \city{Madrid}
  \state{SPAIN}
  \postcode{}
}

\author{Javier P\'{e}rez Trufero}
\email{jptrufero@carto.com}
\affiliation{%
  \institution{CARTO}
  \streetaddress{46 Calle Gran Vía}
  \city{Madrid}
  \state{SPAIN}
  \postcode{}
}

\author{Javier de la Torre}
\email{jatorre@carto.com}
\affiliation{%
  \institution{CARTO}
  \streetaddress{46 Calle Gran Vía}
  \city{Madrid}
  \state{SPAIN}
  \postcode{}
}

\renewcommand{\shortauthors}{Carella, et al.}

\maketitle

\par \textit{Abstract}
\par Fine resolution estimates of demographic and socioeconomic attributes are crucial for planning and policy development. While several efforts have been made to produce fine-scale gridded population estimates, socioeconomic features are typically not available at scales finer than Census units, which may hide local heterogeneity and disparity. In this paper we present a new statistical downscaling approach to derive fine-scale estimates of key socioeconomic attributes. The method leverages demographic and geographical extensive covariates available at multiple scales and additional Census covariates only available at coarse resolution, which are included in the model hierarchically within a ``forward learning'' approach. For each selected socioeconomic variable, a Random Forest model is trained  on the source Census units and then used to generate fine-scale gridded predictions, which are then adjusted to ensure the best possible consistency with the coarser Census data. As a case study, we apply this method to Census data in the United States, downscaling the selected socioeconomic variables available at the block group level, to a grid of $\sim$300 m spatial resolution. The accuracy of the method is assessed at both spatial scales, first computing a pseudo cross-validation coefficient of determination for the predictions at the block group level and then, for extensive variables only, also for the (unadjusted) predicted counts summed by block group. Based on these scores and on the inspection of the downscaled maps, we conclude that our method is able to provide accurate, smoother, and more detailed socioeconomic estimates than the available Census data.

\section{INTRODUCTION}\label{MANUSCRIPT}
\sloppy

From resource allocation, to designing accessibility strategies, and disaster planning, the interest is often on summarizing the characteristics of a location by combining demographic and socioeconomic attributes (e.g. age, population, income, number of household units), geographical features (density of shops, roads, etc.) and, more recently, human mobility patterns from mobile phones \cite{deville2014}. When combining spatial data from disparate sources, policy makers, social and data scientists are frequently faced with the problem of how to best integrate spatial data at multiple scales. Accounting for differences in scale is crucial, because data collected at different scales imply both misaligned geographical regions as well as different data volumes, which, if ignored, can lead to biased analysis \cite{gotway2002}. For example, inferences might depend on the aggregation level or, even for the same scale, on the grouping choice: this source of aggregation and/or zonal bias is a well known issue \cite{openshaw1984,cressie1996}, the so-called Modifiable Areal Unit Problem (MAUP). The problem of transforming the data from one spatial scale to another is known as the change of support problem (COSP), where the support is defined as the size or the volume associated with each data value. Usually, a solution is to transform the data to the finest spatial scale useful for analysis and use that to make inferences \cite{gotway2002}. 

But increasing the spatial resolution of the data (downscaling) is often of interest as such, since data available at a coarse scale may hide local heterogeneity and disparity. This is especially problematic for national population and housing Census data, as the high costs of gathering data through traditional surveying methods make it challenging to study the population characteristics at fine spatial scales. The United Nations has explicitly called for improved availability of ``high-quality, timely, and reliable data disaggregated by income, gender, age, race, ethnicity, migratory status, disability, geographic location, and other characteristics relevant to national contexts'' \cite{assembly2014}, which is crucial to adequately caption sub-national variation in socioeconomic indicators.

To overcome this barrier, international efforts \cite{leyk2019} have been developed to increase the spatial scale of the distribution of human populations. The simplest approach, called dasymetric mapping, consists in redistributing demographic data equally through areal interpolation using covariate data associated with the variable of interest. This is the method used in \cite{tiecke2017}, where for each Census unit the population is distributed equally to settled areas identified from labeled satellite imagery using a Convolutional Neural Network \cite{krizhevsky2012}. While this approach is able to identify settled areas, the estimate of the population for these settled areas within each Census unit will be the same. This represents a limitation when looking at the human population distribution as opposed to settled areas, as for example industrial and commercial areas will be assigned the same population count for the same built-up area. Variations within-settled areas are instead taken into account in the WorldPop dataset (\url{https://www.worldpop.org/}), as described in \cite{stevens2015}. In this case, a Random Forest model \cite{breiman2001} is trained to predict at the Census-unit level the demographic data from remotely-sensed and geographical covariates also available at fine-resolution, and the model weights are then used for the fine-scale redistribution. Both approaches are known as `top-down' population mapping, as opposed to `bottom-up' methods which directly model the population estimates available in small areas from microcensus surveys \cite{wardrop2018}. Compared to the latter, which depend on the availability of  microcensus surveys, `top-down' methods are easier to generalize, although they might rely on out-of-date Census data and the results are usually more difficult to validate.  

`Top-down' methods have been used to derive global estimates of the population (total population, by age and gender categories) but have not been extended, to the best of the authors knowledge, to other socioeconomic attributes from national Census data, restricting the actions and insights of policy makers. The aim of this study is therefore to develop, within the `top-down' framework, a method to derive fine-scale estimates of socioeconomic Census data using geographical and demographic covariates which are also available at finer-resolution (named multiple-scale covariates from now onward). To demonstrate the method, we apply this downscaling framework to national Census data in the United States (US).

The paper is organized as follows: in section \ref{sec_data} we describe the data sources, including the national Census data and the relevant covariates available at multiple scales; in section \ref{sec_methods} we illustrate the method, in section \ref{sec_results} we discuss the results for the US Census data, and finally in section \ref{sec_conc} we draw the conclusions and discuss future work. 

\section{DATA}\label{sec_data}

%\itemize
\begin{enumerate}
\setlength\itemsep{0em}\setlength\parskip{0em}\setlength\topsep{0em}\setlength\partopsep{0em}\setlength\parsep{0em} 
\item{Demographic and socioeconomic Census data. Demographic and socioeconomic data are available for the whole US at the finest-scale at the block group level. The data used in this study are the projected averages for 2019, available from Applied Geographic Solutions (AGS, \url{www.appliedgeographic.com/}). Here, we focus on downscaling a subset of selected variables, which are typically used as a proxy of the socioeconomic status, namely: the number of households and the number of housing units, the median household income, the per capita income, and the median age. As explained in the method section, to improve the downscaling accuracy, the method also takes advantage of other auxiliary socioeconomic attributes (e.g. the total population above 16 years old not in labor force).}

\item{Multiple-scale covariates. The covariates for the `top-down' estimation of socioeconomic Census data should (i) be strongly correlated to each of the variable considered for downscaling, (ii) be available also at a finer scale and (iii) be available everywhere across all areas where the estimation is required. In this study we selected the following covariates, which satisfy these requirements.
\begin{itemize}
     \item[a)] {Geographical covariates}. These include (i) point of interest (POI) data, like the locations of shops and offices, and (ii) road networks as compiled by the US Census (\url{https://www2.Census.gov/}).
     In this study POI data were provided by Pitney Bowes (\url{www.pitneybowes.com}), as proprietary products are expected to be more complete than their open source alternatives, e.g OpenStreetMap \cite{osm}.
     POI data were taken in their Level 2 category by business type (e.g. \textit{MISCELLANEOUS RETAIL}) and re-assigned to lower level categories (e.g. \textit{SHOP}).
     \item[b)] {Demographic covariates.} WorldPop provides global demographic estimates (total and by age and gender) at $\sim$100 m spatial resolution at the Equator. These fine-scale population data are derived downscaling Census data using satellite-derived measures, such as land cover, elevation, lights at night, and geographical data (e.g. water bodies, railways) among others \cite{stevens2015}.
   \end{itemize}
}
\end{enumerate}

Additional information on the data used in this study is available in the Appendix.

\section{METHODS}\label{sec_methods}

This section describes the downscaling method used to transform a random variable $y$ from its source support (\textit{a}), the block group units in this study, to a finer-scale target support (\textit{b}), where this variable is not available or has never been observed.

\subsection{Choice of target support}

The target support was created using quad tree keys \cite{finkel1974}, which provide a common solution to create a spatial index.

The quad tree is a data structure appropriate for storing information that can be retrieved on composite keys (the quadkeys): instead of storing polygons, the Earth is divided into quadrants, each uniquely defined by a string of digits with a length that depends on the level of zoom. A significant advantage of using a quad tree structure is that it captures the hierarchy of parent-children cells: once the grid is provided, the spatial relationships between cells are encoded in the quadkeys, making spatial operations easier (e.g. changing the zoom level). Moreover, by providing a regular grid, a quad tree structure is not locked to any specific boundary type, allowing the same unit anywhere on the globe. 

In this study, the target support was constructed using Google Cloud BigQuery (\url{https://cloud.google.com/bigquery/}) as a quad tree structure with zoom 17, corresponding to a grid of $\sim$300 m by 300 m at the Equator. The choice of zoom 17 is motivated by the need to choose, on the one hand, a support with a coarser spatial resolution than the resolution of the fine-scale demographic covariates, and, on the other hand, with a much finer resolution that the majority of US Census block groups. Given these two conditions, depending on the application, other choices of zoom level are possible.

\subsection{Covariate data preparation}

First, POIs and road networks were processed to derive, for a given geographical unit on the source or target support, the number of POIs by category and the number of road types within that unit.

Secondly, WorldPop demographic data, which are provided as raster files, were converted to vector format and re-projected to the target support by area-interpolation \cite{goodchild1980}, which uses the areas of the intersection as weights.

The data preparation tasks are all performed using Python programming language (version 3.5.3, \url{www.python.org}) and Google Cloud BigQuery.

\subsection{Downscaling framework}

To derive finer-scale estimates for extensive variables (i.e. one whose value for a block can be viewed as a sum of sub-block values, as in the case of population), we can take advantage of extensive covariates available both on the source and the target supports, as those described in the previous section. These extensive covariates, can be summed over each unit on both the source and target supports, effectively driving a model for the change of scale. 

If $y$ is the variable to downscale and $\big\{ x_j \big\}$ a vector of $j = 1,.., J$ covariates, first a model $(f)$ is trained on the source support, where the `ground truth' is available 

\begin{equation}\label{equ:source}
y_{a_m} \sim f \left(\{ x_j^{a_m} \}\right)
\end{equation}

where $a_m$ identifies a unit on the source support $(m = 1,.., M)$. Using this model, predictions are then derived on the target support

\begin{equation}\label{equ:target}
\widehat{y}_{b_l} = f \left(\{  x_j^{b_l} \} \right)
\end{equation}	

where $b_l$ identifies a unit on the target support $(l = 1,.., L)$.

Finally, in order to restore the coherence between the two supports, an `ad hoc' correction is applied on the predictions on the target support, whose magnitude will depend on the ability of the model to learn on the source support and its skill to generalize to a finer spatial support. If \textit{a} and \textit{b} are nested, this correction consists in ``adjusting'' the model predictions on the target support such that when summing over all the unit cells, the corresponding value on the source support is retrieved, i.e.

\begin{equation}\label{equ:mb}
\widehat{y}_{b_l}^{\;\ast} = \widehat{y}_{b_l} \cdot \dfrac{y_{a_m}}{\sum_{b_i \in a_m} \widehat{y}_{b_i}} \end{equation}	

This adjustment represents effectively a way to anchor the model predictions on the target support to the data on the source support, similarly to the dasymetric model-weighted scheme adopted to generate WorldPop estimates \cite{stevens2015}. 

\begin{figure}[h!]
\begin{center}
		\includegraphics[width=0.52
		\columnwidth]{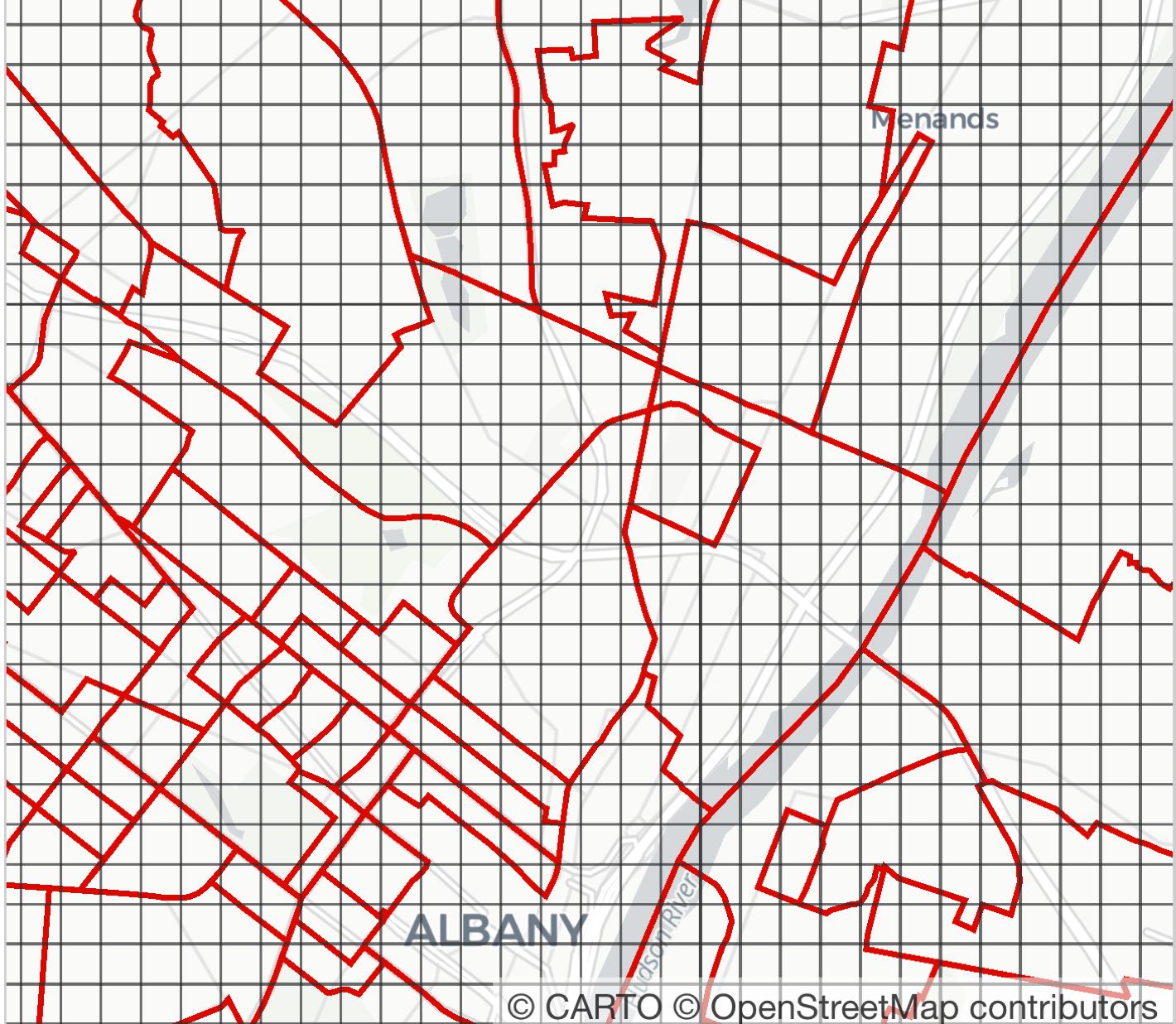}
	\caption{Example of a hybrid support, constructed from the intersection of the source support (block groups, \textit{red polygons}) and the target support (quad trees with zoom 17, \textit{black grid}).}
\label{fig:hybrid_support}
\end{center}
\end{figure}

When the source and the target support are not nested, for the prediction task, one must instead resort to a hybrid support (\textit{c}), obtained by the spatial intersection of the source and target support. This is illustrated in figure \ref{fig:hybrid_support}, where the source support is represented by US block groups and the target support is constructed as a regular grid using quad trees with zoom 17.

The predictions on the target support are finally obtained by summing over all the unit cells on the hybrid support corresponding to a cell on the target support after applying the \textit{ad hoc} adjustment discussed above (c.f. eq. \ref{equ:mb}).

For non-extensive variables (e.g. the median household income), it is still possible to use a similar approach based on multiple-scale extensive covariates, training the model on the source support and predicting on the hybrid support. However, in this case the \textit{ad hoc} adjustment is interpreted as a weighting scheme based on the model trained on the source support for ``redistributing'' the data rather than a correction for the change of scale. In this case, it is also necessary to specify the relevant transformation, as multiple options are possible (e.g. the arithmetic mean, the median, etc.), which has to be applied both when adjusting the predictions on the hybrid support and when transforming them from the hybrid to the target support. In the following, this transformation for non-extensive variables was chosen as the arithmetic mean.

Finally, we note that while the method presented here requires complete response data on the source support, an extension for missing data is also possible. In this case, the model would be trained on complete units only and then used for prediction everywhere on the hybrid support, with the difference that for those units on the hybrid support that correspond to units on the source support with missing response data, the adjustment detailed in Eq. \ref{equ:mb} would not be possible. 

\subsection{Choice of statistical model}

First, in order to reduce the model dimension, only a subset of the available covariates were selected (a.k.a. feature selection). This selection step was implemented by using a Generalized Linear Model (GLM) with a Least Absolute Shrinkage and Selection Operator (LASSO) regularization \cite{tibshirani1994}. Adding a LASSO regularization allows in fact to select the most relevant covariates for a given response variable and model: by decreasing the size of the regularization parameter, more covariates are included up to the full model (or the largest identified model when there are more covariates than observations, which is not the case here). The conditional distribution of the response variable that best fits the data was selected by comparing the cross-validation mean squared error for different family types (Gaussian and Poisson). For the chosen distribution, the optimal regularization parameter was then selected such that the cross-validation mean squared error is within one-standard deviation of its minimum.

Secondly, in order improve the model skills, a ``forward learning'' (FL) approach was adopted. While the distribution of human population, POIs, and road networks might be expected to be relevant predictors for certain dependent variables, like the number of households, a much smaller predictive power is observed for income-related variables (c.f section \ref{sec_results}). Within this framework, a second step is added where extra covariates are hierarchically included in the model. These extra covariates are chosen from the pool of available Census variables (e.g. the total population above 16 years old not in labor force) provided that their model depending on the multiple-scale covariates only has higher accuracy than the current model. 

\begin{figure*}[h!]
\begin{center}
		\includegraphics[width=1\columnwidth]{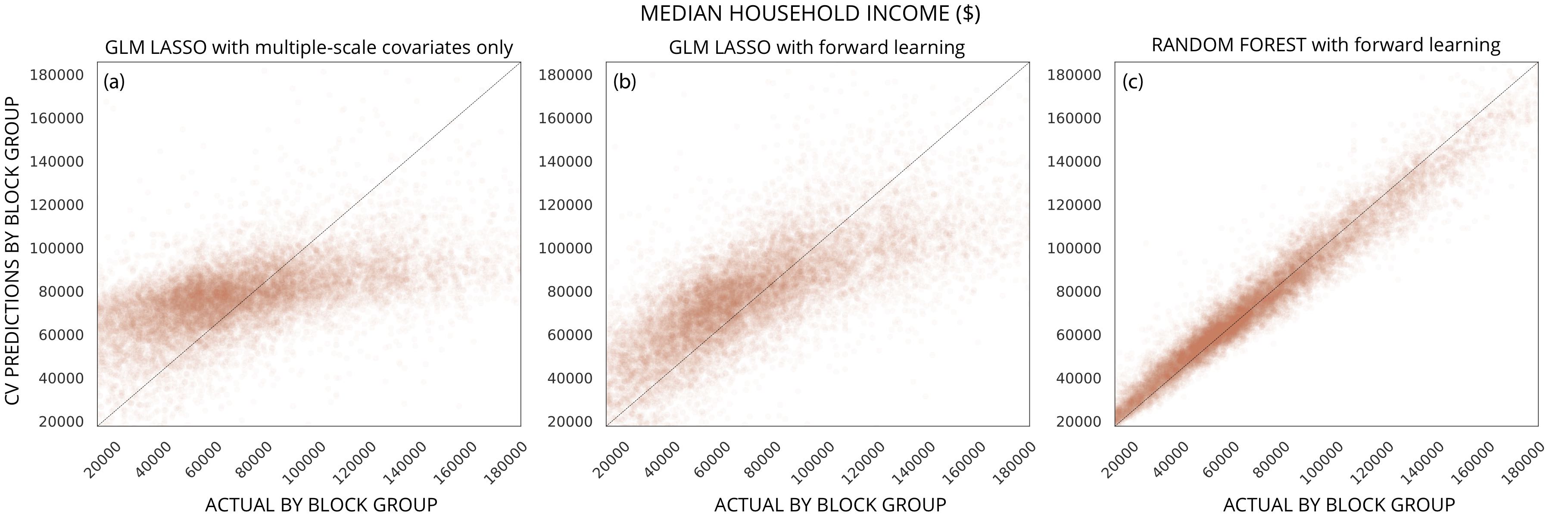}
	\caption{Cross validation predictions vs. actual by block group for the median household income in NY state. (a) with the GLM LASSO model with multiple-scale covariates only, (b) with the GLM LASSO model with forward learning, and (c) with the Random Forest model with forward learning. The same 5-folds were used for all three models.}
	\Description{}
\label{fig:forward_learning}
\end{center}
\end{figure*}

Model accuracy is measured computing, for the cross-validation predictions ($\widehat{y_i}$), the pseudo-$R^2$ score as

\begin{equation}
\label{equ:score}
\text{pseudo-}R^2 = 1- \dfrac{\sum_i (y_i-\widehat{y_i})^2}{\sum_i (y_i-\overline{y})^2} 
\end{equation}

where $\overline{y}$ is the mean of the response variable $y$.

Because the extra covariates might include any other of the selected variables for downscaling, this step could lead to a set of nested models, which we will refer to as ``forward learning'' (FL). For example, by training on the source support, based on the multiple-scale covariates only, a model for $y_1$ and then training a second model for $y_2$ with both the multiple-scale features and $y_1$ as covariates, we ``learn'' on the source support to transfer the features that are predictive of the variations for $y_1$ to $y_2$. Under the assumption of independence of scale, this learning framework can then be adopted to generate hierarchically, first for $y_1$ and then for $y_2$, fine-scale estimates on the target support. This framework allows to account for model misspecification for those variables for which the true data-generating process is too complex to be feasibly captured given the available data and limited tools. As for the model including fine-scale available covariates only, in order to keep the model dimension as low as possible, the LASSO regularization was adopted also in this second step. 

The pseudo-$R^2$ scores for the model using the multiple-scale covariates only are used to sort the prediction order on the target support: first, predictions are generated for those variables whose model depends only on the multiple-scale covariates, which are then in turn included as covariates for the remaining variables, with the last predicted variable being that not included as covariate in any model. 

\begin{figure*}[h!]
\begin{center}
		\includegraphics[width=1\columnwidth]{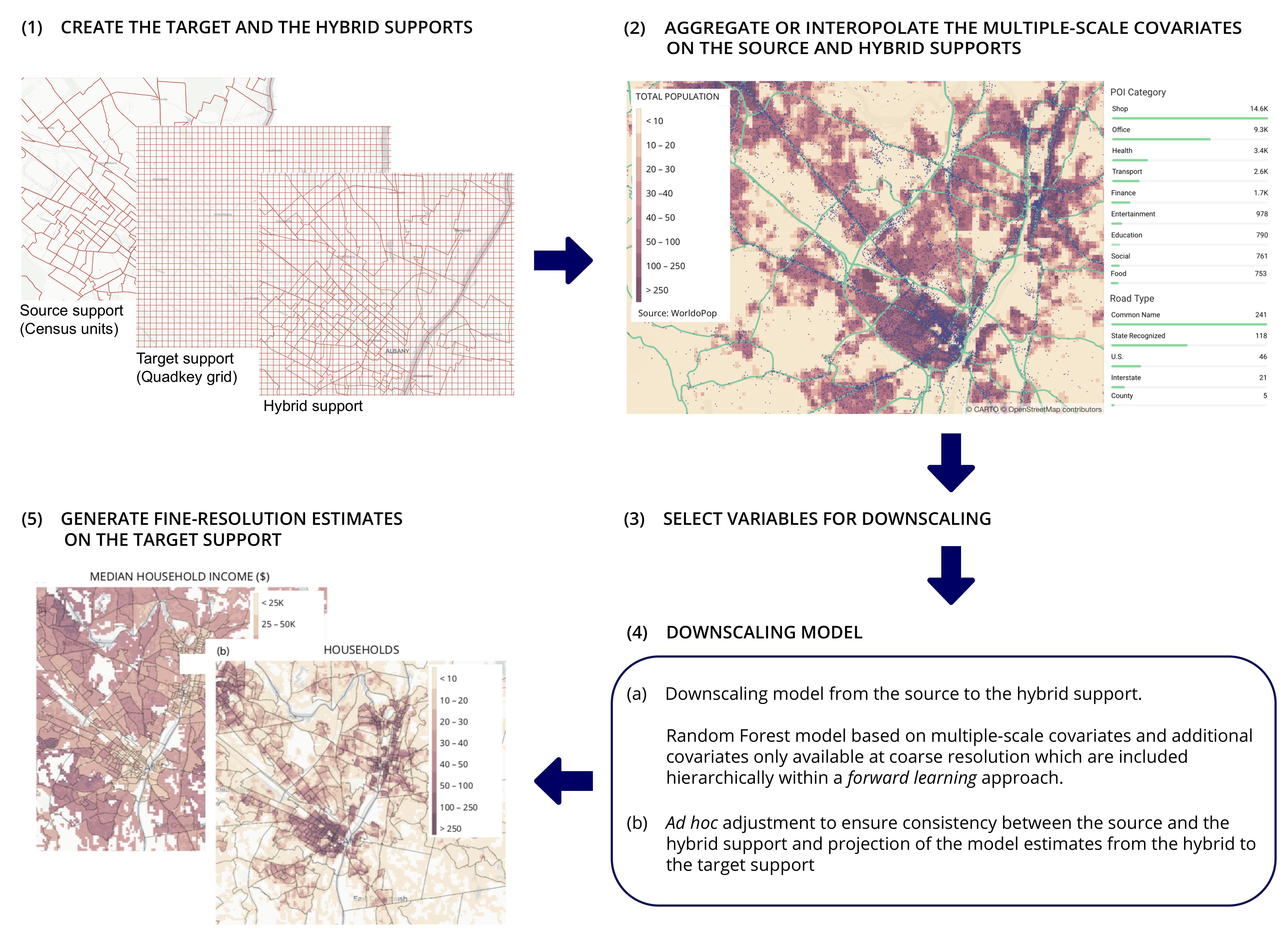}
	\caption{Diagram illustrating the steps in the downscaling framework.(1) Create the target and hybrid supports; (2) Aggregate or interpolate the multiple-scale covariates on the source and hybrid supports; (3) Select variables for downscaling; (4) Downscaling model; (5) Generate fine-resolution estimates on the target support.}
\label{fig:flowchart}
\end{center}
\end{figure*}

In order to account for non-linear dependencies of the response variable on the covariates selected on the regularized forward learning step, the final estimates were obtained with Random Forest \cite{breiman2001}, a non-parametric ensemble tree model. The model hyper-parameters were chosen as those yielding the lowest cross-validation mean squared error. While hyper-parameters selection is necessary to prevent over-fitting, the RF algorithm has the advantage of relying on fewer tuning parameters compared to other methods, making it a good candidate in this semi-automated framework, as already observed in \cite{stevens2015}. On the other hand, as similar ensemble tree methods or artificial neural networks, RF predictions are limited to the range of values spanned by the training observations, hindering their applicability in some cases. In the current downscaling framework though, given an appropriate choice of training units (from largely inhabited, to rural, suburban and urban units), such limitation does not apply, as the range of the multiple-scale covariates on the target support does not extend beyond the bounds of their range on the source support.

Figure \ref{fig:flowchart} shows a diagram illustrating the steps in the downscaling framework.

The downscaling model was applied separately for data in the state of New York (NY) and Missouri (MO). While New York City has one of the highest population density and Gross Domestic Product (GDP) of the country, Central and upstate New York are regions in economic and social decline \cite{NYEmployment}, with towns and smaller cities gradually emptying out due to movement of industrial plants out of the state. On the other hand, Missouri comprises large farmland and rural areas, with two major cities accounting for the overall state population. While out of the scope of this study, in order to better analyse the model sensitivity to the training set, a comparison between the results obtained by training each individual state separately as opposed to training the model with data from the whole US would be required and should be explored in future studies.

Feature selection using LASSO regularization was performed using the statistical environment R version 3.6.1 \cite{Rcore} and the glmnet (2.0.18) package \cite{friedman2010}, while model estimation and prediction for the Random Forest model were implemented using Python programming language and the hyperopt-sklearn (0.19.2) package \cite{bergstra2015}. 

\subsection{Accuracy assessment}

The accuracy of the downscaling model was first assessed computing the pseudo-$R^2$ score for the cross-validation predictions at the block group level, where the `ground truth' is known. 

On the other hand, testing the model accuracy for the predictions on the target support is more challenging, as fine-scale socioeconomic data are not available throughout the US. For extensive variables however, the correlation between the block group values and the unadjusted predicted counts on the quad tree grid summed by block group could serve as an approximate metric for assessing the model dependence on the change of scale in the absence of `ground truth' fine-scale estimates.  A low correlation would indicate that the model is highly dependent on the spatial scale, and therefore might suggest potentially biased downscaled estimates. 

\section{RESULTS}\label{sec_results}

Figure \ref{fig:forward_learning} shows for the median household income the scatter plot of the cross-validation predictions on the source support (at the block group level) vs. the Census data. The results for the GLM model using multiple-scale covariates only are reported in panel (a), while those for the forward learning GLM and RF models, where also other census variables are included as covariates, are shown in panel (b) and (c) respectively. To make a fair comparison between the different models, five folds with the same random split were selected for all three models. While the choice of the cross-validation folds might influence the model accuracy, especially for spatial data which tend to be characterized by correlated errors, this was found not to be the case here, with the pseudo-$R^2$ scores only varying by few percent points (not shown).

As can be seen from these plots and looking at the pseudo-$R^2$ scores in table \ref{tab:r2_scores_source}, the forward learning approach improves notably the model accuracy for income-related variables (the median household income and the per capita income), more than doubling the scores both for NY and MO. Comparing the performance of the FL-GLM and FL-RF model, as expected, a non-parametric model only improves the accuracy for those variables for which the score for the linear model is low. This is also well illustrated comparing panel (b) and (c) in figure \ref{fig:forward_learning} for the median household income, which shows the improvement in accuracy obtained with the RF model, despite a minor over-prediction (under-prediction) bias at low (high) values can still be observed.

\begin{table}[ht]
  \caption{Cross-validation pseudo-$R^2$ scores for the GLM-LASSO, GLM-LASSO with FL, and RF with FL models. Results are listed both for NY and MO. The same 5-folds were used for all three models.}
    \footnotesize
    \begin{tabular}{ccccccc} %  
    \toprule
    \multirow{2}{*}{Variable name} & \multicolumn{6}{c}{\centering 5-fold CV pseudo-$R^2$ score} \\ 
     \cmidrule{2-7}
     & \multicolumn{2}{p{0.09\textwidth}}{\centering{GLM-LASSO}} & \multicolumn{2}{p{0.09\textwidth}}{\centering{FL GLM-LASSO}} & \multicolumn{2}{p{0.09\textwidth}}{\centering{FL RF}}   \\
      \cmidrule{2-3} \cmidrule{4-5} \cmidrule{6-7}
    & \centering{NY} & \centering{MO} & \centering{NY} &  \centering{MO}& \centering{NY} & MO \\
    \midrule
    Median household income  & 0.26 & 0.25 & 0.50 & 0.54 & 0.63 & 0.66\\
    \\
    Per capita income  & 0.18 & 0.21 & 0.45 & 0.70 & 0.55 & 0.70\\
    \\
    Median age & 0.65 & 0.69 & 0.67 & 0.80 & 0.91  & 0.84\\
     \\
    Households & 0.87 & 0.94 & 0.96 & 0.97 & 0.90 & 0.97\\
    \\
    Housing units & 0.79 & 0.86 & 0.98 & 0.98 & 0.93 & 0.98\\
    \bottomrule
    \end{tabular}
\label{tab:r2_scores_source}
\end{table}

Additionally the comparison, for the extensive variables, between the block group values and the grid-level predictions summed by block group, showed a good correlation, with an $R^2$ score of 0.92 (NY) and 0.78 (MO) for the households variable and of 0.98 (NY) and 0.57 (MO) for the housing units variable respectively. While this result suggests that overall the model behaves well even at finer scales, the inspection of the summed residuals, showed for the state of Missouri, for both the households and the housing units variables, typically positive, larger residual values concentrated in the suburban areas (not shown).

\begin{figure*}[h!]
\begin{center}
		\includegraphics[width=1\columnwidth]{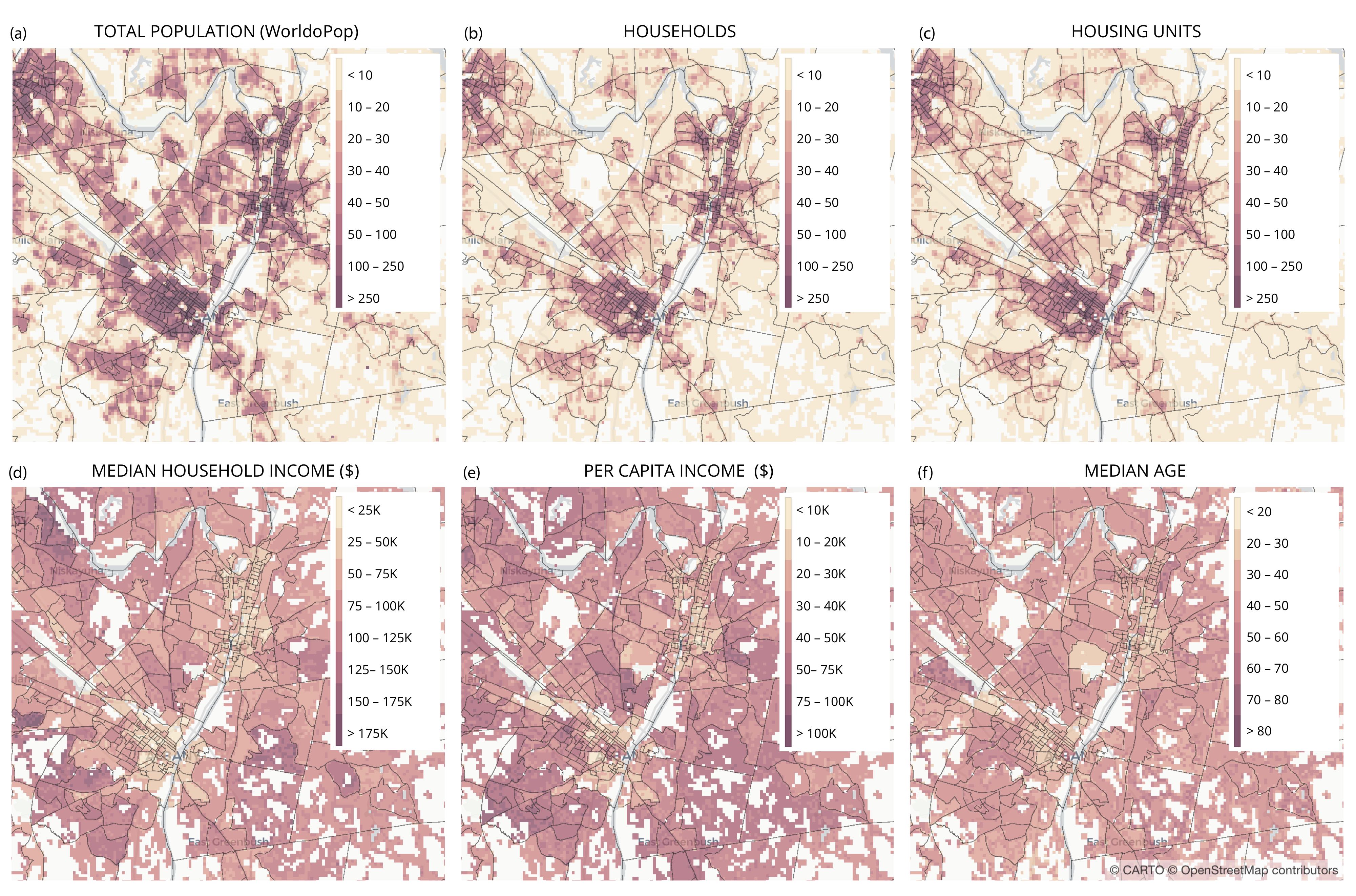}
	\caption{(a) Total population from the WorldPop dataset and downscaled (b) households, (c) housing units, (d) median household income, (e) per capita income, and (f) median age near Albany, NY USA. Only grid cells with a total population of at least one are shown.}
\label{fig:downscaled_maps}
\end{center}
\end{figure*}

\begin{figure*}[h!]
\begin{center}
		\includegraphics[width=0.5\columnwidth]{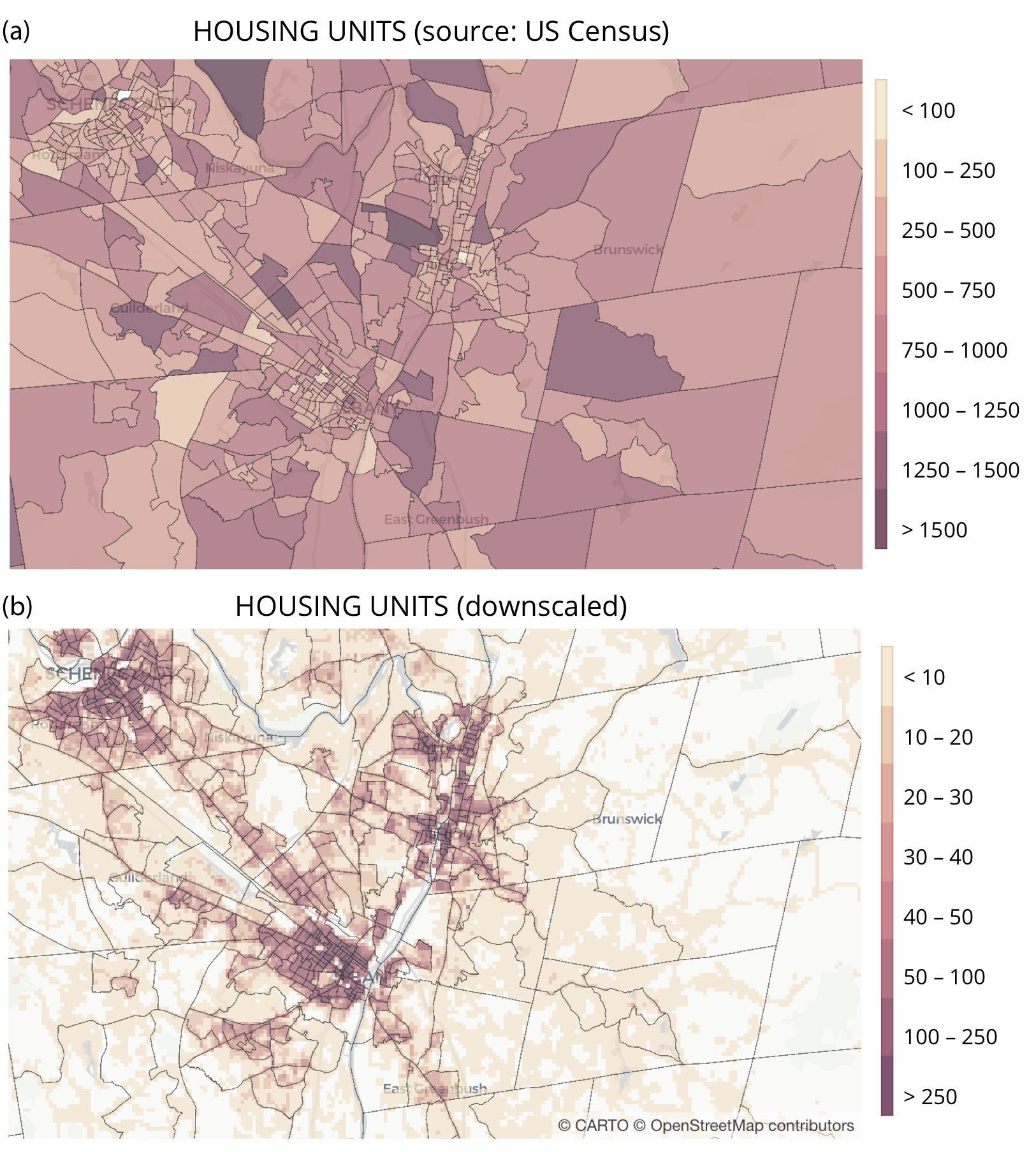}
	\caption{Housing units: Census data by block group (a) downscaled (b).}
\label{fig:downscaled_raw_maps}
\end{center}
\end{figure*}

Finally, figure \ref{fig:downscaled_maps} shows the (adjusted) downscaled outputs for the selected variables near Albany, NY. As these maps illustrate, the downscaled estimates are characterized by a large variability within each Census unit, which is driven by the multiple-scale covariates. For the same reason, near the borders, a smoother, more continuous redistribution compared to the coarser block group level values is observed, as also illustrate by \ref{fig:downscaled_raw_maps}.

\section{DISCUSSION AND CONCLUSIONS}\label{sec_conc}

In this study, we presented a novel approach to derive fine-scale estimates of key socioeconomic attributes from Census data that otherwise would be only available at coarse scale, providing increased spatial details for policy analysis and support. 

The method leverages demographic and geographical extensive covariates available at multiple scales and additional Census covariates only available at coarse resolution, which are included in the model hierarchically within a ``forward learning'' approach. For each selected socioeconomic attribute, a Random Forest model is trained  on the source Census units and then used to generate fine-scale gridded predictions, which are then adjusted to ensure the best possible consistency with the coarser Census data. This method was applied in the states of New York and Missouri to downscale, on a target grid of about 300 m x 300 m at the Equator, block group level data for the number of households, the number of housing units, the median household income, the per capita income, and the median age. 

First, the model accuracy was assessed at the block group level, by comparing the cross-validation predictions with the actual data. As our results show, the method yields good predictions for all the selected variables. A net improvement in the model accuracy at the block group level is obtained for income-related variables using the forward learning approach compared to a model based on the multiple-scale covariates only.

Despite the unavailability of fine-scale estimates, for extensive variables only (here the households and the housing units variables), the accuracy of the downscaled predictions was also assessed, to some extent, at the target grid level, by comparing the actual and (unadjusted) predicted counts summed by block group. Large differences would indicate that the model is strongly dependent on scale, and despite producing good predictions at the block group level does not generalize well to finer spatial scales. A good correlation between the summed and the actual counts was observed for both the households and the housing units variables, although lower correlation scores were found for Missouri, which might be explained by differences in the average block group size (larger for Missouri than NY state). While these results indicate that overall the models for the households and the housing units produce predictions that are consistent with Census data, larger values for the summed residuals were observed for both variables for suburban areas in Missouri (not shown). Given these observed spatially correlated residuals, although Random Forest represents a flexible modeling approach capable of dealing with collinearity and non-linearity in covariate dependence, for further developments of the method, we believe that a model explicitly accounting for spatially correlated errors might produce more accurate results. 

Given the availability of Census data at coarser scales (e.g. Census tracts), in order to assess the model accuracy at different scales, it should be noted that an ``upscaling'' approach could also be tested. In this framework, the model is trained on this coarser support and then the predictions are derived at the block group level, allowing point-to-point comparisons with the `ground truth'. However, as existing literature on the MAUP \cite{openshaw1984} suggests, this approach does not anyhow guarantee that the same results would be obtained for a different change of scale (e.g. from block groups to a finer quad tree structure), limiting the strength of any conclusions.
 
Despite these considerations, by leveraging a model of demographic and geographical covariates available at multiple scales, the downscaling method presented here allows for smoother and more detailed socioeconomic estimates. The resulting maps could be of great value for policy and planning initiatives, from the generation of development indicators to improve resource allocation, accessibility, and disaster planning amongst others. This framework also offers a general methodology that is not limited to socioeconomic data: provided that relevant multiple-scale covariates are available, our downscaling model can be generalized to other data sources for which the access to fine-scale estimates is limited.

% Acknowledgments
\begin{acks}
This research is made possible by CARTO. Special thanks to Stuart Lynn and Wenfei Xu for stimulating discussions and comments related to this work.
\end{acks}

% Appendices
\newpage
\appendix

\section*{APPENDIX}

\setcounter{table}{0}
\renewcommand{\thetable}{A\arabic{table}}

\begin{table*}[h]
	\caption{Demographic and socioeconomic variables in the US Census.}
    \footnotesize
		\begin{tabular}{*{2}{l}}
		    \toprule 
		    \multicolumn{1}{c}{\textbf{Variable name}} &  \multicolumn{1}{c}{\textbf{Description}}   \\
            \midrule
             POPCY  & Population \\
             AGECY  & Population by age \\
             SEXCY  & Population by sex \\
             RCHCY  & Race \\
             MARCY  & Civil status \\
             EDUCY  & Education level \\
             HHDCY  & Households and Household types \\
             VPHCY  & Number of vehicles \\
             DWLCY  & Housing units \\
             HUSEX  & Units in structure \\
             INCCY  & Income \\
             HINCY  & Household income by income and age range \\
             LBFCY  & Employment \\
             UNECYRATE & Employment rate \\
             LNIEX  & Linguistic types \\
             HOOEXMED  & Median Value of Owner Occupied Housing Units \\
             RNTEXMED  & Median Cash Rent \\
 		    \bottomrule
		\end{tabular}
\label{tab:mbi_var}
\end{table*}

\begin{table*}[h!]
	\caption{POI categories.}
    \scriptsize
		\begin{tabular}[x]{*{2}{l}}
		    \toprule 
		    \multicolumn{1}{c}{\textbf{POI category}} &  \multicolumn{1}{c}{\textbf{Group (Level 2 POI category by business type)}}   \\
            \midrule
                ACCOMMODATION & HOTELS, ROOMING HOUSES, CAMPS, AND OTHER LODGING PLACES\\
                EDUCATION & EDUCATIONAL SERVICES\\
                ENTERTAINMENT & TOURISM \\ & SPORTS \\ & MUSEUMS, ART GALLERIES AND BOTANICAL AND ZOOLOGICAL GARDENS \\ & MOTION PICTURES \\ & LEISURE \\ & AMUSEMENT AND RECREATION SERVICES\\
                FINANCE & PUBLIC FINANCE, TAXATION AND MONETARY POLICY \\ & NON-DEPOSITORY CREDIT INSTITUTIONS \\ & DEPOSITORY INSTITUTIONS\\
                FOOD & EATING AND DRINKING PLACES\\
                HEALTH & HEALTH SERVICES\\
                OFFICE & UNITED STATES POSTAL SERVICE \\ & SERVICES, NEC \\ & SECURITY AND COMMODITY BROKERS, DEALERS, EXCHANGES AND SERVICES \\ & SECURITY AND COMMODITY BROKERS, DEALERS, EXCHANGES AND SERVICES \\ & REAL ESTATE \\ & NATIONAL SECURITY AND INTERNATIONAL AFFAIRS \\ & MEMBERSHIP ORGANIZATIONS \\ & LEGAL SERVICES \\ & JUSTICE, PUBLIC ORDER AND SAFETY \\ & INSURANCE CARRIERS \\ & INSURANCE AGENTS, BROKERS AND SERVICE \\ & HOLDING AND OTHER INVESTMENT OFFICES \\ & GOVERNMENT AND PUBLIC SERVICES \\ & EXECUTIVE, LEGISLATIVE AND GENERAL GOVERNMENT, EXCEPT FINANCE \\ & SPECIAL TRADE CONTRACTORS \\ & CONSTRUCTION - GENERAL CONTRACTORS AND OPERATIVE BUILDERS \\ & ADMINISTRATION OF HUMAN RESOURCE PROGRAMS \\ & ADMINISTRATION OF ENVIRONMENTAL QUALITY AND HOUSING PROGRAMS \\ &COMMUNICATIONS \\ &ELECTRIC, GAS AND SANITARY SERVICES \\ &HOLDING AND OTHER INVESTMENT OFFICES \\ &EXECUTIVE, LEGISLATIVE AND GENERAL GOVERNMENT, EXCEPT FINANCE \\ &ADMINISTRATION OF HUMAN RESOURCE PROGRAMS \\ &ADMINISTRATION OF ECONOMIC PROGRAMS \\ &ENGINEERING, ACCOUNTING, RESEARCH, AND MANAGEMENT SERVICES\\
                SOCIAL & SOCIAL SERVICES\\
                SHOP & FOOD STORES \\ &FOOD AND KINDRED PRODUCTS \\ &SHOPPING \\ & PERSONAL SERVICES \\ & MISCELLANEOUS RETAIL \\ & MISCELLANEOUS REPAIR SERVICES \\ & HOME FURNITURE, FURNISHINGS AND EQUIPMENT STORES \\ &FURNITURE AND FIXTURES \\ & GENERAL MERCHANDISE STORES \\ & BUSINESS SERVICES \\ & BUILDING MATERIALS, HARDWARE, GARDEN SUPPLIES AND MOBILE HOMES \\ & AUTOMOTIVE REPAIR, SERVICES AND PARKING \\ & AUTOMOTIVE DEALERS AND GASOLINE SERVICE STATIONS \\ & APPAREL AND ACCESSORY STORES \\ &APPAREL, FINISHED PRODUCTS FROM FABRICS AND SIMILAR MATERIALS \\ &WHOLESALE TRADE - NON-DURABLE GOODS \\ &LEATHER AND LEATHER PRODUCTS \\ &ELECTRONIC AND OTHER ELECTRICAL EQUIPMENT, AND COMPONENTS \\ &MEASURING, PHOTOGRAPHIC, MEDICAL, AND OPTICAL GOODS, AND CLOCKS\\
                TRANSPORT & TRANSPORTATION SERVICES \\ & TRANSPORT \\ & RAILROAD TRANSPORTATION \\ & LOCAL AND SUBURBAN TRANSIT AND INTERURBAN HIGHWAY TRANSPORTATION\\
 		    \bottomrule
		\end{tabular}
\label{tab:poi_cat}
\end{table*}

\clearpage

\bibliographystyle{acm}
\bibliography{bibliography}

\end{document}